\documentclass[prl,%
 reprint,
 amsmath,amssymb,
 aps,
]{revtex4-2}
\usepackage{textcomp} 
\newcommand{\qdist}[1]{\ifmmode\langle#1\rangle\else\textlangle#1\textrangle\fi}

\usepackage{amsmath}
\usepackage{empheq}
\usepackage{tikz}
\usepackage{amsthm}
\usepackage{float}
\usepackage{amssymb}
\usepackage{comment}
\usepackage{graphicx}
\usepackage{dcolumn}
\usepackage{bm}
\usepackage{braket}
\usepackage{physics}
\usepackage{hyperref}
\usepackage{cleveref}

\usepackage{quantikz}
\usepackage[mathlines]{lineno}
\usepackage{amsfonts}
\usepackage[bb=boondox]{mathalfa}
\usepackage{braket}
\usepackage{dsfont}
\usepackage{xcolor}
\usepackage[colorinlistoftodos,prependcaption,textsize=tiny]{todonotes} 

\begin{document}

\preprint{APS/123-QED}

\title{A Tractable Protocol for Detection-Loophole-Free Bell Tests over Long Distances}

%

%
%

%

%
%



\author{Yazeed K. Alwehaibi}
\email{y.alwehaibi23@imperial.ac.uk}
\affiliation{Blackett Laboratory, Department of Physics, Imperial College London, Prince Consort Rd, London, SW7 2AZ, United Kingdom}
\affiliation{Centre for Quantum Engineering, Science and Technology (QuEST), Imperial College London, Prince Consort Rd, London, SW7 2AZ, United Kingdom}

\author{Ewan Mer}
\affiliation{Blackett Laboratory, Department of Physics, Imperial College London, Prince Consort Rd, London, SW7 2AZ, United Kingdom}
\affiliation{Centre for Quantum Engineering, Science and Technology (QuEST), Imperial College London, Prince Consort Rd, London, SW7 2AZ, United Kingdom}

\author{Gerard J. Machado}
\affiliation{Blackett Laboratory, Department of Physics, Imperial College London, Prince Consort Rd, London, SW7 2AZ, United Kingdom}
\affiliation{Centre for Quantum Engineering, Science and Technology (QuEST), Imperial College London, Prince Consort Rd, London, SW7 2AZ, United Kingdom}

\author{Shang Yu\footnote{These authors contributed equally to this work}\footnote{shang.yu@imperial.ac.uk}}
\affiliation{Blackett Laboratory, Department of Physics, Imperial College London, Prince Consort Rd, London, SW7 2AZ, United Kingdom}
\affiliation{Centre for Quantum Engineering, Science and Technology (QuEST), Imperial College London, Prince Consort Rd, London, SW7 2AZ, United Kingdom}

\author{Ian A. Walmsley}
\affiliation{Blackett Laboratory, Department of Physics, Imperial College London, Prince Consort Rd, London, SW7 2AZ, United Kingdom}
\affiliation{Centre for Quantum Engineering, Science and Technology (QuEST), Imperial College London, Prince Consort Rd, London, SW7 2AZ, United Kingdom}

\author{Raj B. Patel}
\email{raj.patel1@imperial.ac.uk}
\affiliation{Blackett Laboratory, Department of Physics, Imperial College London, Prince Consort Rd, London, SW7 2AZ, United Kingdom}
\affiliation{Centre for Quantum Engineering, Science and Technology (QuEST), Imperial College London, Prince Consort Rd, London, SW7 2AZ, United Kingdom}



\date{\today}

\begin{abstract}

Certifying genuine nonclassical correlations over long distances is crucial for device-independent (DI) quantum information protocols. However, in photonic platforms this remains technologically challenging due to photon loss, which opens the detection-loophole, rendering violations increasingly difficult for less efficient detectors. A well-known strategy to mitigate this involves using non-maximally entangled states, which Eberhard showed can tolerate lower detection efficiencies. However, existing proposals and demonstrations have been limited to short distances, as their success rates scale linearly with channel transmittance. Here, we propose a protocol to herald a tunable entangled state between distant users that achieves a post-selection-free Bell inequality violation at the Eberhard limit. We identify the loss independence of the vacuum component amplitude of the prepared state as the source of this enhancement. Notably, our scheme requires only quantum interference at a central station, followed by the detection of a single photon, preserving the optimal square-root scaling with channel transmittance. Our approach provides greater loss-tolerance in entanglement distribution, bringing long-distance DI applications closer to practical implementation.
\end{abstract}

\maketitle


A cornerstone of device-independent (DI) quantum information protocols—where neither the source nor the measurement devices are trusted—is the violation of a loophole-free Bell inequality for appropriate variables \cite{barrett2005no,acin2007device,pironio2009device,pironio2010random}. Among the various loopholes in Bell tests,  the detection-loophole is perhaps the most challenging to close as it requires a total end-to-end throughput---hereby referred to as the detection efficiency---exceeding 82.6\% to ensure unconditional violation when using maximally entangled states \cite{christensen2013detection, giustina2013bell}. Photonics is the only viable platform for distributing entanglement over appreciable distances. However, photon loss can drastically hinder the end-to-end efficiency of a communication channel, even with high-efficiency detectors, thus making the prospect of closing the detection-loophole even more daunting in practice.

To date, such tests have only been demonstrated over a few hundred meters \cite{shalm2015strong, hensen2015loophole, storz2023loophole, giustina2015}---the distance required for space-like separation between two parties, Alice and Bob. Recently, there has been growing interest in \textit{event-ready} protocols based on single-photon path-entanglement (SPPE) as a promising approach to mitigate photon loss \cite{vivoli2015comparing, mycroft2023proposal, steffinlongo2024long, caspar2020heralded}. These protocols are attractive because the detection efficiency scales more favourable \cite{lucamarini2018overcoming} when used to herald entanglement between remote parties. Specifically, the success rate scales proportional to $\sqrt{\eta_C}$, where $\eta_C$ denotes the total transmittance of the channel connecting the users. Moreover, these entanglement-based protocols offer the potential for performing long-distance loophole-free Bell tests, and thus enable DI operation—a key milestone for secure quantum communication \cite{gisin2010proposal,vivoli2015comparing,mycroft2023proposal,steffinlongo2024long,caspar2020heralded}.

Eberhard showed that the detection efficiency threshold for violating a Bell inequality can be reduced to 2/3 ($\approx 66.7\%$) by using non-maximally entangled states \cite{eberhard1993background}. In photonic implementations, this approach has been employed to close the detection loophole in two-photon polarisation entanglement experiments \cite{christensen2013detection,giustina2013bell,giustina2015,shalm2015strong}. For photon-number entanglement (up to single-photon), the Eberhard limit can be reached using displacement-based measurements, provided the state exhibits perfect photon-number correlation—that is, both modes are either in the vacuum or contain a single photon simultaneously \cite{mcdermott2022eberhard}. However, existing long-distance entanglement distribution protocols typically generate the anticorrelated SPPE state, which cannot reach the Eberhard limit and instead requires detection efficiencies exceeding 82.6\% for post-selection-free Bell tests \cite{vivoli2015comparing,mycroft2023proposal}.

The fact that these photon-number path-entangled (PNPE) states exhibit different loss tolerance is nuanced and surprising, given that both are Bell-like states and are related by a local unitary transformation. This contrast becomes even more striking when compared to the case of two-photon polarisation entanglement, where both correlated and anticorrelated Bell-like states can, as expected, achieve the Eberhard limit. To the best of our knowledge, a clear explanation for this disparity in the photon-number setting has not yet been established. 
\begin{figure*}[t!]
\includegraphics[width=1.0\linewidth]{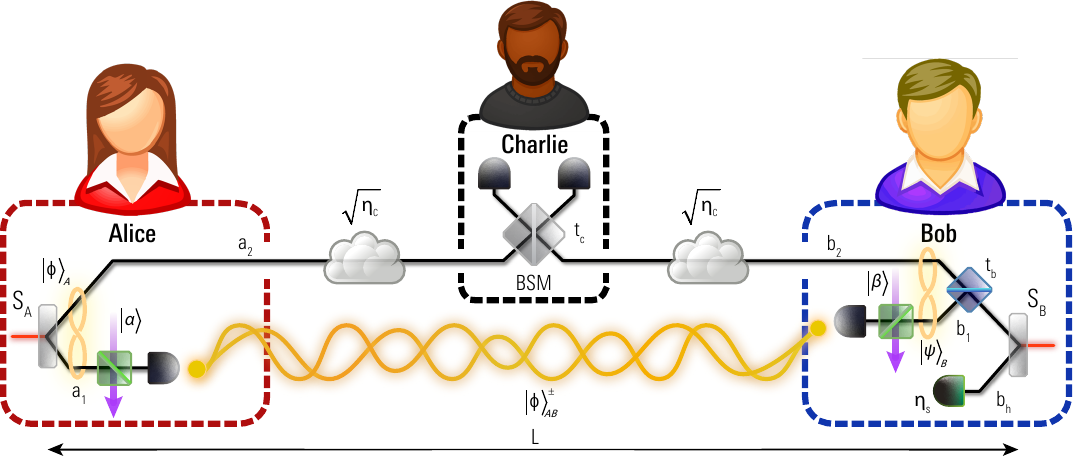}
\caption{\label{fig:layout}An illustration of our protocol for the heralded distribution of photon-number path-entanglement (PNPE). Alice and Bob, separated by a distance $L$, generate entangled states using sources $\text{S}_A$ and $\text{S}_B$, respectively. The modes $a_2$ and $b_2$ are directed to a middle station (Charlie), which performs a Bell-state measurement (BSM), thereby heralding entanglement between the modes $a_1$ and $b_1$. The photons travelling to Charlie experience a channel transmittance of $\sqrt{\eta_C}$, and a successful BSM occurs if exactly one photon arrives. The modes $a_1$ and $b_1$ are then interfered with weak local oscillators ($\ket{\alpha}$, $\ket{\beta}$) to implement displacement-based measurements for the Bell test. In this illustration, the sources are based on SPDC: Alice prepares a two-mode squeezed vacuum state, while Bob generates a heralded single-photon path-entangled state (SPPE).}
\end{figure*}

In this Letter, we introduce a protocol for producing the required state to reduce the detection efficiency threshold down to the Eberhard limit, between distant parties, while maintaining the favourable square-root scaling with channel transmittance. Our approach promises to overcome the longstanding trade-off between closing the detection-loophole and enabling long-distance operation. We evaluate the performance of our protocol and compare it with existing schemes based on SPPE and polarisation entanglement. Finally, we present a theoretical analysis that explains the origin of the contrasting loss tolerances observed among different PNPE states in the context of Bell inequality violations.

The notion of heralding entanglement between Alice and Bob, based on a single-photon detection at a central station, has been previously reported, for example, in Ref.~\cite{mycroft2023proposal}, where both parties use two-mode squeezed vacuum (TMSV) states, and Ref.~\cite{steffinlongo2024long}, where SPPE states are used. In both cases, the heralded state is SPPE, which is highly sensitive to loss.

\textit{Protocol.---} We propose an event-ready protocol, illustrated in Fig.~\ref{fig:layout}, in which Alice prepares a TMSV state while Bob generates an SPPE state. This asymmetry in the initial states enables heralding of the target entangled state that enables this significantly improved performance. A practical realisation uses type-II spontaneous parametric down-conversion (SPDC) in the low-gain regime, which directly produces the required TMSV state for Alice
\begin{equation}
\ket{\phi}_A=\dfrac{1}{\sqrt{1+\lambda^2}}\left(\ket{0,0}_{a_1,a_2}+\lambda e^{i\varphi_a}\ket{1,1}_{a_1,a_2}\right),
\label{eq:Alices_real_InitialState}
\end{equation}
where $\lambda = \tanh g$,  $g$ is the squeezing parameter, $\varphi_a$ is the pump phase. Since $\lambda \ll 1$, we ignore higher-order photon-number contributions for this illustration, although they are included in the numerical simulation. Bob prepares his SPPE state by heralding on a single-photon detection and sending the other photon through a beam splitter with transmittance $t_b$, yielding
\begin{equation}
\ket{\psi}_B=\sqrt{t_b}\,\ket{0,1}_{b_1,b_2}\,+\sqrt{1-t_b}e^{i\varphi_b}\ket{1,0}_{b_1,b_2},
\label{eq:Bobs_real_InitialState}
\end{equation}
where $\varphi_b$ is the path phase difference. Since Bob prepares his state by heralding on a single-photon detection with efficiency $\eta_s$, the state $\ket{\psi}_B$ is generated with a success probability $P_s=\eta_s\lambda^2/(1+\lambda^2)$. We ignore the probability that a heralding detector response yields more than one photon in the state, which is reasonable for extremely low parametric gains.  

Alice and Bob are separated by a distance $L$, as shown in Fig.~\ref{fig:layout}, and each sends one mode of their bipartite state ($a_1$ and $b_1$) to a central station (Charlie) located midway between them, at a distance $L/2$. 

Charlie performs a partial Bell state measurement (BSM) using a beamsplitter (BS), with transmittance $t_c$, followed by click, or photon-number-resolving detection. The transmittance of the direct channel between Alice and Bob is given by $\eta_C=10^{-\gamma L/10}$, where $\gamma$ is the attenuation coefficient, which implies a transmittance of $\sqrt{\eta_C}$ to the midpoint station. Now, if the modes are indistinguishable after entering Charlie's BS, it becomes impossible to determine the origin of a single detection event at Charlie's detectors. As a result, the shared state between Alice and Bob is projected onto one of the desired (unnormalised) states
\begin{equation}
\ket{\phi}^{\pm}_{AB}=\sqrt{t_b}\ket{0,0}_{a_1,b_1}\pm \lambda\sqrt{1-t_b}e^{i\varphi}\ket{1,1}_{a_1,b_1},
\label{eq:sharedstate}
\end{equation}
where $\varphi=\varphi_a+\varphi_b$, and the exact state depends on which detector clicks at Charlie (see Supplemental Material Sec. I \cite{supplemental} for the derivation). The degree of entanglement in these states can be continuously tuned—from a product state to a maximally entangled state—by adjusting the amplitudes of the initial states. While the beam splitter at Charlie’s station was assumed to be balanced in Eq.~(\ref{eq:sharedstate}), its transmittance can also serve as an additional degree of freedom to tune the entangled state.

The final heralded state, including symmetric losses in the transmitted modes (i.e., $\eta_{a_2}=\eta_{b_2}=\eta_H$), is
\begin{equation}
    \begin{aligned}
        \rho^{(\pm)}_{AB}&=\mathcal{N}^{-1}\Big( [\ket{\phi}\bra{\phi}]^{\pm}_{AB} +\varepsilon[\ket{1,0}\bra{1,0}]_{a_1,b_1}\Big),
    \end{aligned}
    \label{eq:heralded_state_symm_loss}
\end{equation}
where $\mathcal{N}^{-1}$ is the normalisation factor, $\varepsilon=2\lambda^2 t_b(1-\eta_H)$, and $\eta_H = \eta^c_D \sqrt{\eta_C}$, with $\eta^c_D$ denoting Charlie’s detection efficiency (see Supplemental Material Sec. II \cite{supplemental} for the full derivation, including the treatment of
asymmetric losses). We observe that Eq.~(\ref{eq:heralded_state_symm_loss}) is equivalent to the desired state in Eq.~(\ref{eq:sharedstate}), up to an additional noise term ($\ketbra{1,0}{1,0}$), which can be neglected when $\varepsilon \ll 1$. In this limit, which can be plausibly reached, the shared state is effectively pure, independent of channel losses. The overall success probability is, however, naturally reduced, as the heralding efficiency appears in the state normalisation, trading off the degree of shared entanglement with the distribution rate. 

By tuning $g$ of each source and $t_b$, one can control the degree of entanglement in the shared state. For instance, a maximally entangled state between Alice and Bob is achieved when the two sources have the same $g$ and the condition $\sqrt{t_b} = \lambda \sqrt{1 - t_b}$ is satisfied. To further quantify the degree of entanglement, we compute the log-negativity of the shared state and examine its dependence on the tuning parameters; see Supplemental Material Sec. III \cite{supplemental}. This tunability highlights the versatility of our protocol, enabling it to be adapted to diverse experimental constraints and optimised for various levels of detection efficiency to maximise the Bell violation as will be shown later.



\begin{figure}
    \centering
\includegraphics[width=1\linewidth]{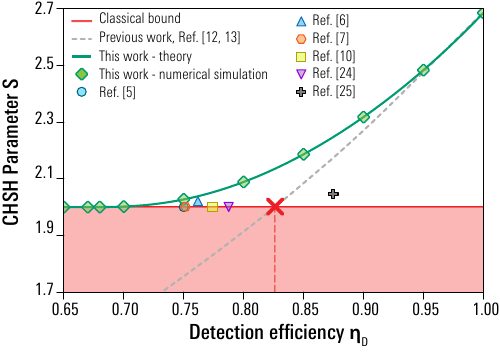}
       \caption{Optimal CHSH violations as a function of symmetric detection efficiency $\eta_D$. The solid green curve shows the theoretical prediction for our protocol, with numerical simulations that account for higher-order photon-number contributions (green diamonds) in excellent agreement. The gray dashed line corresponds to previous photon-number-based protocols, which require detection efficiencies above 82.6\% (red cross and dropline) to observe a violation. Coloured markers denote detection-loophole-free violations from leading polarisation-based experiments. While those can operate at the Eberhard limit (66.7\%), the observed violations are marginal and their distance scaling is limited. In contrast, our protocol achieves significantly stronger violations down to the Eberhard limit, demonstrating enhanced robustness to loss and improved overall performance.}
	       \label{fig:optimal_settings}
    \end{figure}

\begin{figure}
    \centering
\includegraphics[width=1\linewidth]{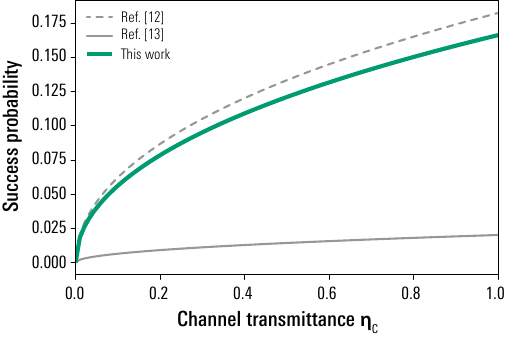}
    \caption{Success probability of performing the BSM at Charlie and heralding entanglement between Alice and Bob. The squeezing parameter $g$ is fixed at 0.33. For Ref.~ \cite{steffinlongo2024long} and for this work, the BS transmittance is set to 0.1 and 0.8, respectively. The parameters are chosen to remain consistent with the operating assumptions and optimal regimes of each protocol.}
\label{fig:performance}
\end{figure}

\begin{table*}
\caption{\label{tab:comparison_table}Comparison of heralded entangled states and key performance metrics for the protocols discussed in the main text.}
\begin{ruledtabular}
\begin{tabular}{lllll}
\#&Protocol & \hspace{1cm} Heralded State& Scaling& Required Detection Efficiency \\ \hline
 1&SPPE-based&$\ket{\psi}_{AB}=\frac{1}{\sqrt{2}}(\ket{0,1}_{a,b}+e^{i\varphi}\ket{1,0}_{a,b})$ &$\mathcal{O}(\sqrt{\eta_C})$&\hspace{1.3cm}$\geq82.6\%$ \\
 2&Polarisation&$\ket{\Psi}_{AB}=\epsilon\ket{H,H}_{a,b}+\sqrt{1-\epsilon^2} e^{i\varphi}\ket{V,V}_{a,b}$&$ \mathcal{O}(\eta_C)$&\hspace{1.2cm} $\geq66.67\%$\\
3& This work&$\ket{\phi}_{AB}=\epsilon\ket{0,0}_{a,b}+\sqrt{1-\epsilon^2} e^{i\varphi}\ket{1,1}_{a,b}$&$\mathcal{O}(\sqrt{\eta_C})$&\hspace{1.2cm} $\geq66.67\%$\\
\end{tabular}
\end{ruledtabular}
\end{table*}
\textit{Protocol Performance.---} Once the entangled state is shared, Alice and Bob interfere their respective modes with weak coherent states $\ket{\alpha}$ and $\ket{\beta}$ to perform displacement-based measurements. By setting the amplitudes and phases of these local oscillators, they can perform a post-selection-free Bell test to verify the nonlocality of the shared state \cite{banaszek1999testing,kuzmich2000violation,donati2014} (see Supplemental Material Sec. IV \cite{supplemental} for the full derivation of the CHSH and CH inequalities). 

Figure~\ref{fig:optimal_settings} presents the central result of this work, showing a comparison across different protocols, each optimised for maximum CHSH inequality violation as a function of detection efficiency. Our protocol achieves a violation with total detection efficiency as low as 67\%, reaching the Eberhard limit, while existing photon-number-based protocols~\cite{mycroft2023proposal,steffinlongo2024long} require efficiencies above 82.6\%\footnote{The proposal in Ref.~\cite{steffinlongo2024long} reduces the required efficiency to 76.7\% using a more complex setup that incorporates a squeezer after the displacement operation. This goes beyond the standard displacement-based measurements considered here and still remains above the Eberhard limit.}. The figure also includes results from state-of-the-art polarisation-entanglement experiments \cite{giustina2013bell,christensen2013detection,shalm2015strong,giustina2015,li2018test,liu2022toward}. While these setups can also reach the Eberhard limit, the observed violations are notably small. Indeed, it has been shown that the maximum achievable post-selection-free CHSH violation drops from approximately 2.82 to 2.35 for polarization-based schemes using a single SPDC source~\cite{caprara2015challenging}, and even lower values in heralded versions of such schemes~\cite{seshadreesan2016progress,tsujimoto2020heralded}. This is compared to a maximum of 2.69 for PNPE protocols. Remarkably, our protocol surpasses existing schemes not only in the low-efficiency regime but it also achieves stronger violations across a wide range of detection efficiencies, establishing a new benchmark for the achievable violation in post-selection-free nonlocality tests using SPDC sources. The excellent agreement between our theoretical model and numerical simulations, which account for higher-order photon-number contributions from the TMSV, further validates the assumptions made in our analysis.

Since our protocol and those in Refs.~\cite{mycroft2023proposal,steffinlongo2024long} rely on quantum interference at Charlie’s station followed by the detection of a single photon to perform a successful BSM, they exhibit a success probability that scales as $\sqrt{\eta_C}$. The probability of performing a successful BSM at Charlie—and thereby heralding entanglement between Alice and Bob—using this approach can be calculated as
\begin{equation}
 P_{\text{succ}}=\sqrt{\eta_C}(P_A(1)+P_B(1))-2\eta_C P_A(1)P_B(1).
\label{eq:psuccess}
\end{equation}
Here, $\sqrt{\eta_C}$ denotes the channel transmittance from Alice or Bob to Charlie, and $P_x(1)$ is the probability that party $x$ prepares a state with one photon in the transmitted mode (see Supplemental Material Sec. V \cite{supplemental} for full details). Equation~(\ref{eq:psuccess}) clearly shows that the success probability of these protocols scales as $\mathcal{O}(\sqrt{\eta_C})$, provided that $P_A(1)P_B(1) \ll 1$. It is straightforward to use Eq.~(\ref{eq:psuccess}) together with Alice and Bob's states to calculate the success probability for each protocol. Fig.~\ref{fig:performance} shows these probabilities as a function of the channel transmittance when using SPDC sources with equal squeezing parameters. While all protocols exhibit the expected square-root dependence on transmittance, their success probabilities differ. The protocol using two TMSV states \cite{mycroft2023proposal} achieves the highest success rate due to the relatively high probability of producing single photons in the transmitted modes. In contrast, the protocol using two heralded single-photon entangled states \cite{steffinlongo2024long} has a lower success rate. Since our scheme uses a combination of both source types, its success probability lies in between the two. It is worth noting that with our protocol, one can trade off some degree of Bell violation to further increase the success probability (see Supplemental Material Sec. V \cite{supplemental}).

The event-ready polarisation-based protocols with SPDC sources suffer from two significant shortcomings. First, they rely on quantum interference at the heralding station with the detection of two photons, which causes the success probability to scale linearly with the channel transmittance \cite{kolodynski2020device}. Second, due to the possibility of both photons coming from the same source, the fidelity of the heralded entangled state is reduced — only one-third of the successful heralding events correspond to the desired case where one photon originates from each source\cite{seshadreesan2016progress,tsujimoto2020heralded}.

\Cref{tab:comparison_table} provides a brief summary of the differences in the heralded states and performance metrics across the protocols we compare \footnote{Assuming an ideal polarisation-entangled state. Note that the anticorrelated Bell state exhibits the same performance.}. Notably, our protocol is the only one that simultaneously achieves both the optimal square-root scaling and the Eberhard limit.


\textit{Realization and Applications.---} Given that our protocol can operate at the Eberhard limit without post-selection, except in the state preparation stage, it can be readily implemented using current photon sources and detectors. In fact, total detection efficiencies exceeding 67\%—typically in the range of 73–78\%—have already been demonstrated by several experimental groups \cite{giustina2013bell,christensen2013detection,shalm2015strong,giustina2015,li2018test,liu2022toward}, making our scheme well-suited for near-term, long-distance, all-photonic implementations of DI quantum information tasks such as DI quantum key distribution (DI-QKD), randomness extraction, and certification\cite{mycroft2023proposal,steffinlongo2024long,acin2016certified,vivoli2015comparing,pironio2010random}.


As concrete use cases, we evaluate two key quantities relevant to device-independent applications: the extractable randomness per round and the Holevo bound. The rate of certifiable randomness $R$ can be lower bounded by the min-entropy via $R \geq r H_{\min}$, where $r$ is the experimental repetition rate, typically limited by the overall detection rate. The second quantity, $\chi_{\max}$, is the Holevo bound, which quantifies the maximum information that an eavesdropper could gain in a DI-QKD protocol with one-way classical communication from Bob to Alice. Both quantities can be expressed as functions of the observed CHSH value $S$ \cite{vivoli2015comparing,pironio2009device}
\begin{subequations}
\begin{align}
H_{\text{min}}&=1-\log_2\Big(1+\sqrt{2-S^2/4}\;\Big)\label{eq:minEntropy}, \\
\chi_{\text{max}}&=h\big(\dfrac{1+\sqrt{S^2/4-1}}{2}\Big)\label{eq:holevo},
\end{align}
\label{eq:min_entropyANDholevo}
\end{subequations}
where $h(x)$ is the binary entropy function.

\begin{figure}
	\centering
	\includegraphics[width=1\linewidth]{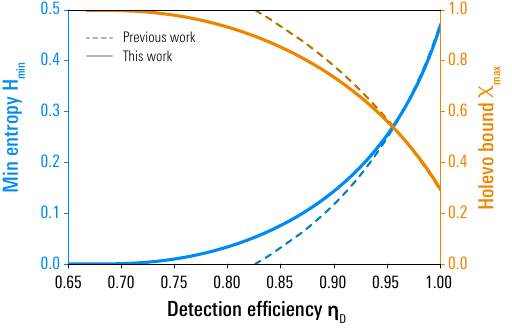}
    \caption{Min-entropy $\text{H}_{\min}$ (left axis, blue) and Holevo bound $\chi_{\max}$ (right axis, orange) as functions of detection efficiency $\eta_D$ for the protocols compared. Solid lines correspond to the present work; dashed lines represent previous approaches (Refs.~\cite{mycroft2023proposal,steffinlongo2024long}). Increasing $\text{H}_{\min}$ indicates higher certifiable randomness, while decreasing $\chi_{\max}$ reflects reduced eavesdropper information in DI-QKD.}
	\label{fig:MinEntropyANDHolevo}
\end{figure}

Given that polarisation-based schemes yield only marginal violations of Bell inequalities, we restrict our comparison to PNPE-based protocols. Figure~\ref{fig:MinEntropyANDHolevo} presents both the min-entropy $H_{\min}$ and the Holevo bound $\chi_{\max}$ as functions of the detection efficiency $\eta_D$. As expected, higher CHSH violations yield larger values of $H_{\min}$, indicating greater certifiable randomness, and smaller values of $\chi_{\max}$, reflecting stronger privacy guarantees and a higher potential secret key rate in DI-QKD. Notably, below the threshold $\eta_D = 82.6\%$, only our protocol provides positive rates for both randomness and key generation. Moreover, it consistently outperforms previous methods across a broad range of detection efficiencies (up to $\sim 95\%$), beyond which all protocols converge. These results demonstrate the enhanced robustness and practical viability of our scheme for both DI randomness generation and QKD.

\textit{Discussion.---} The question arises as to why the state $\ket{\phi} = c_0\ket{0,0} + c_1\ket{1,1}$ can reach the Eberhard limit, with our measurement scheme, while the seemingly equivalent state $\ket{\psi} = c_0\ket{0,1} + c_1\ket{1,0}$ cannot.
This distinction is, in one sense, counterintuitive: both states are Bell-like and related by a local unitary, suggesting that they could yield the same violation in principle \cite{wildfeuer2007strong}. In the case of two-photon polarisation entanglement, such a difference does not arise; these states can, in principle, achieve the Eberhard limit. It is also reasonable to expect this behaviour in dual-rail encoding of other degrees of freedom, such as time-bin and frequency. 

The key to understanding this tolerance is to note that loss does not affect the vacuum, but only the components of the state that contain photons. In the state in $\ket{\psi}$, both amplitudes are reduced when there is loss present (and a vacuum component introduced), whereas in the state $\ket{\phi}$, only the second amplitude is reduced. 




Perhaps most remarkably, the minimum detection efficiency required to observe a violation with $\ket{\phi}$ approaches the Eberhard limit for two-qubit entanglement. A full quantitative analysis of the CHSH inequality for each state in the presence of loss is presented in Supplemental Material Sec. VI \cite{supplemental},

In conclusion, we have proposed an event-ready protocol that heralds a tunable PNPE state upon single-photon detection. Our scheme combines the favourable twin-field-like scaling, $\mathcal{O}(\sqrt{\eta_C})$, with resilience to detection inefficiencies. Our analysis shows that this robustness arises from the coherent superposition of the biphoton state with vacuum, enabling Bell violations even under high-loss conditions. Notably, our protocol achieves a violation at detection efficiencies as low as \(66.7\%\), corresponding to the Eberhard limit for two-qubit entanglement. Given that current experimental platforms already exceed this threshold, our results open a practical route toward near-term, long-distance, all-photonic implementations of device-independent quantum technologies. Moreover, the heralded state can be tuned to the canonical Bell state $\ket{\phi}$, making it broadly useful for a wide range of photonic quantum information tasks beyond Bell tests.
\begin{acknowledgments}
\textit{Acknowledgements.---} The authors would like to thank Magdalena Stobinska, Morteza Moradi, Maryam Afsary, and Krister Shalm for valuable discussions. 

This work was funded by UK Research and Innovation Future Leaders Fellowship (project MR/W011794/1) and the National Company of Telecommunications and Information Security (NTIS) and partly funded by National Research Council of Canada (project QSP 062-2), EU Horizon 2020 Marie Sklodowska-Curie Innovation Training Network (project no. 956071, `AppQInfo'), UK Research and Innovation Guarantee Postdoctoral Fellowship (project EP/Y029631/1).
\end{acknowledgments}

\bibliographystyle{apsrev4-2}
\bibliography{apssamp}
\onecolumngrid
\newpage

\section*{Supplemental Material}
\renewcommand{\theequation}{S\arabic{equation}}
\setcounter{equation}{0}
\renewcommand{\thefigure}{S\arabic{figure}}
\setcounter{figure}{0}
\setcounter{table}{0}
\renewcommand{\thetable}{S\Roman{table}}

\section*{\label{app:derivation of sharedstate}I. Ideal Heralded State}
The idealised states for evaluating the protocol for preparing entanglement between Alice and Bob are:
\begin{subequations}
\begin{align}
\ket{\phi}_A&=a_{0}\ket{0,0}_{a_1,a_2}+a_1\ket{1,1}_{a_1,a_2}\label{eq:AlicesInitialState}\\
\ket{\psi}_B&=b_{0}\,\ket{0,1}_{b_1,b_2}\,+b_1\ket{1,0}_{b_1,b_2}\label{eq:BobssInitialState}.
\end{align}
\label{eq:app_initialstates}
\end{subequations}
where $|a_0|^2+|a_1|^2=1$ and $|b_0|^2+|b_1|^2=1$. Then, the initial joint state is $\ket{\Psi}_i=\ket{\phi}_A\otimes\ket{\psi}_B$
\begin{equation}
\begin{aligned}
\ket{\Psi}_i&=\ket{\phi}_A\otimes\ket{\psi}_B\\
&=a_0b_0\ket{0,0}_{a_1,a_2}\ket{0,1}_{b_1,b_2}+a_0b_1\ket{0,0}_{a_1,a_2}\ket{1,0}_{b_1,b_2}\\
&+a_1b_0\ket{1,1}_{a_1,a_2}\ket{0,1}_{b_1,b_2}+a_1b_1\ket{1,1}_{a_1,a_2}\ket{1,0}_{b_1,b_2}.
\end{aligned}
\label{eq:approx_in_state}
\end{equation}

After the beamsplitter interaction at Charlie, the state becomes
\begin{equation}
\begin{aligned}
\ket{\Psi}_{ABC}&=BS_{a_2,b_2}\ket{\Psi}_i\\&=\ket{1,0}_{a_2,b_2}[a_0b_0\sqrt{1-t_c}\ket{0,0}_{a_1,b_1}+ a_1b_1\sqrt{t_c}\ket{1,1}_{a_1,b_1}]\\
&-\ket{0,1}_{a_2,b_2}[a_0b_0\sqrt{t_c}\ket{0,0}_{a_1,b_1}- a_1b_1\sqrt{1-t_c}\ket{1,1}_{a_1,b_1}].
\end{aligned}
\label{eq:state_after_CBS}
\end{equation}

where $BS_{a_2,b_2}$ is the unitary matrix of the beamsplitter interaction at Charlie, and $t_c$ is its transmittance. Eq.~(\ref{eq:state_after_CBS}) shows that a measurement at Charlie will project the shared state between Alice and Bob into the desired state. The exact form depends on which detector clicked at Charlie. Specifically, if: 

\begin{enumerate}
    \item Charlie's $D_{c_2}$ detector clicks, then 
    \begin{equation}
        \ket{\phi}^+_{AB}=a_0b_0\sqrt{1-t_c}\ket{0,0}_{a_1,b_1}+ a_1b_1\sqrt{t_c}\ket{1,1}_{a_1,b_1}.
    \end{equation}

    \item Charlie's $D_{c_1}$ detector clicks, then 
    \begin{equation}
        \ket{\phi}^-_{AB}=a_0b_0\sqrt{t_c}\ket{0,0}_{a_1,b_1}- a_1b_1\sqrt{1-t_c}\ket{1,1}_{a_1,b_1}.
    \end{equation}
\end{enumerate}

Using the proposed realisation, $a_0=\dfrac{1}{\sqrt{1+\lambda^2}}$, $a_1=\dfrac{\lambda}{\sqrt{1+\lambda^2}}$, $b_0=\sqrt{t_b}$, and $b_1=\sqrt{1-t_b}$.

\section*{II. General Heralded State}
\label{app:heralded_state}
Here, we derive the heralded state when using the proposed realisation of the protocol, by including the channel losses in the transmitted modes ($a_2$ and $b_2$), and Charlie's detectors efficiency, although we retain the maximum photon number in any mode at one. We start from the state $\ket{\Psi}_i=\ket{\phi}_A\otimes\ket{\psi}_B$, were $\ket{\phi}_A$ and $\ket{\psi}_B$ are defined in Eq.~(1) and Eq.~(2) in the main text, respectively. We model the total loss between Charlie and Alice (Bob) with a beamsplitter of transmittance $\eta_{a_2}$ ($\eta_{b_2}$). This includes the channel losses and Charlie's detector efficiencies, as the loss associated with Charlie's detector commutes with his Bell-state measurement beamsplitter. Then, the initial state after the transmitted modes of Alice and Bob interact with the environment modes $e_a$ and $e_b$, respectively, becomes
\begin{equation}
    \begin{aligned}
    \ket{\Psi}_l
= \dfrac{1}{\sqrt{1+\lambda_a^2}}&\Bigg[\ket{0,0}_{e_a,e_b} \Big( \sqrt{t_b}\sqrt{\eta_{b_2}} \ket{0,0}_{a_1,b_1} \ket{0,1}_{a_2,b_2} + \sqrt{1-t_b}e^{i\varphi_b} \ket{0,1}_{a_1,b_1}  \ket{0,0}_{a_2,b_2} 
 \\&+ \lambda_a e^{i\varphi_a}\sqrt{t_b} \sqrt{\eta_{a_2}} \sqrt{\eta_{b_2}} \ket{1,0}_{a_1,b_1} \ket{1,1}_{a_2,b_2}+ \lambda_a e^{i\varphi_a}\sqrt{1-t_b}e^{i\varphi_b} \sqrt{\eta_{a_2}} \ket{1,1}_{a_1,b_1} \ket{1,0}_{a_2,b_2} \Big)\\
 &+ \ket{0,1}_{e_a,e_b}\Big( \sqrt{t_b} \sqrt{1-\eta_{b_2}} \ket{0,0}_{a_1,b_1} \ket{0,0}_{a_2,b_2} + \lambda_a e^{i\varphi_a}\sqrt{t_b} \sqrt{\eta_{a_2}} \sqrt{1-\eta_{b_2}} \ket{1,0}_{a_1,b_1} \ket{1,0}_{a_2,b_2} \Big) \\
&+\ket{1,0}_{e_a,e_b}\Big( \lambda_a e^{i\varphi_a}\sqrt{t_b} \sqrt{1-\eta_{a_2}} \sqrt{\eta_{b_2}} \ket{1,0}_{a_1,b_1} \ket{0,1}_{a_2,b_2} + \lambda_a e^{i\varphi_a}\sqrt{1-t_b}e^{i\varphi_b} \sqrt{1-\eta_{a_2}} \ket{1,1}_{a_1,b_1} \ket{0,0}_{a_2,b_2} \Big) \\
&+\ket{1,1}_{e_a,e_b}\Big( \lambda_a e^{i\varphi_a}\sqrt{t_b} \sqrt{1-\eta_{a_2}} \sqrt{1-\eta_{b_2}} \ket{1,0}_{a_1,b_1} \ket{0,0}_{a_2,b_2} \Big)\Bigg]
    \end{aligned}
    \label{eq:initial_state_with_idlers_loss}
\end{equation}
The state entering Charlie's station is found by tracing out the environment modes. That is $\widetilde{\rho}=\text{Tr}[\ket{\Psi}_l\bra{\Psi}_l]_{e_a,e_b}$. Charlie conducts the projective measurement described by

\begin{equation}
    \Pi^{(\pm)}_C=\dfrac{1}{2}(\ketbra{0,1}{0,1}\pm\ketbra{0,1}{1,0}\pm\ketbra{1,0}{0,1}+\ketbra{1,0}{1,0})_{a_2,b_2}
    \label{eq:Charlie_projector}
\end{equation}

Finally, the heralded state between Alice and Bob is given by
\begin{equation}
    \begin{aligned}
        \rho^{(\pm)}_{AB}&=\mathcal{N}^{-1}\text{Tr}[\Pi^{(\pm)}_C\widetilde{\rho}\Pi^{\dagger(\pm)}_C]_{a_2,b_2}\\
        &=\mathcal{N}^{-1}\Big[ t_b \eta_{b_2} \ketbra{0,0}{0,0}\pm  \lambda\sqrt{\eta_{a_2}\eta_{b_2}t_b(1-t_b)} e^{-i (\varphi_a + \varphi_b)}\ketbra{0,0}{1,1} \\
        &\pm \lambda\sqrt{\eta_{a_2}\eta_{b_2}t_b(1-t_b)} e^{i (\varphi_a + \varphi_b)}\ketbra{1,1}{0,0}+\lambda^2 (1-t_b) \eta_{a_2}\ketbra{1,1}{1,1}+\lambda^2 t_b (\eta_{a_2}+\eta_{b_2}-2\eta_{a_2}\eta_{b_2})\ketbra{1,0}{1,0}\Big]_{a_1,b_1}
    \end{aligned}
    \label{eq:heralded_state}
\end{equation}
where $\mathcal{N}^{-1}=\text{Tr}[\Pi^{(\pm)}_C\widetilde{\rho}\Pi^{\dagger(\pm)}_C]$ is the normalization factor. In the case of symmetric loss and equal detector efficiencies, $\eta_{a_1}=\eta_{b_1}=\eta_H$, this becomes

\begin{equation}
    \begin{aligned}
        \rho^{(\pm)}_{AB}&=\eta_H\mathcal{N}^{-1}\Big[ t_b  \ketbra{0,0}{0,0}\pm  \lambda\sqrt{t_b(1-t_b)} e^{-i (\varphi_a + \varphi_b)}\ketbra{0,0}{1,1} \\
        &\pm \lambda\sqrt{t_b(1-t_b)} e^{i (\varphi_a + \varphi_b)}\ketbra{1,1}{0,0}+\lambda^2 (1-t_b) \ketbra{1,1}{1,1}+2\lambda^2 t_b(1-\eta_H)\ketbra{1,0}{1,0}\Big]_{a_1,b_1}\\
        &=\mathcal{N'}^{-1}\Big( [\ket{\phi}\bra{\phi}]^{\pm}_{AB} +2\lambda^2 t_b(1-\eta_H)[\ket{1,0}\bra{1,0}]_{a_1,b_1}\Big)
    \end{aligned}
    \label{eq:heralded_state_symm_loss2}
\end{equation}

where $\ket{\phi}^{\pm}_{AB}$ is Eq.~(3) and $\mathcal{N'}^{-1}=\eta_H\mathcal{N}^{-1}$.

\section*{\label{app:LogNegativity}III. Log Negativity}
To further quantify the degree of entanglement, we compute the log‑negativity of the shared state, defined as
$\mathcal{E}_N(\rho)\;=\;\log_2\bigl\|\rho^{T}\bigr\|_1$,
where $\rho^{T}$ is the partially transposed density matrix and $\|\cdot\|_1$ denotes the trace norm.
In the limit $\lambda^2 t_b\ll1$, the heralded density matrix becomes

\begin{equation}
    \begin{aligned}
        \rho^{(\pm)}_{AB}&=\mathcal{N'}^{-1} \Big[ t_b  \ketbra{0,0}{0,0}\pm  \lambda\sqrt{t_b(1-t_b)} e^{-i (\varphi_a + \varphi_b)}\ketbra{0,0}{1,1} \\
        &\pm \lambda\sqrt{t_b(1-t_b)} e^{i (\varphi_a + \varphi_b)}\ketbra{1,1}{0,0}+\lambda^2 (1-t_b) \ketbra{1,1}{1,1}\Big]_{a_1,b_1}
    \end{aligned}
    \label{eq:heralded_state_symm_loss2}
\end{equation}

Thus, the partial transpose with respect to B is

\begin{equation}
    \begin{aligned}
        \rho^{T_B}_{\pm}&=\mathcal{N'}^{-1} \Big[ t_b  \ketbra{0,0}{0,0}\pm  \lambda\sqrt{t_b(1-t_b)} e^{-i (\varphi_a + \varphi_b)}\ketbra{0,1}{1,0} \\
        &\pm \lambda\sqrt{t_b(1-t_b)} e^{i (\varphi_a + \varphi_b)}\ketbra{1,0}{0,1}+\lambda^2 (1-t_b) \ketbra{1,1}{1,1}\Big]_{a_1,b_1}
    \end{aligned}
    \label{eq:heralded_state_symm_loss2}
\end{equation}

The eigenvalues of this density matrix can be easily read as
$\mathcal{N'}^{-1}\bigl\{\,t_b,\;\lambda^2(1-t_b),\;+\lambda\sqrt{t_b(1-t_b)},\;-\lambda\sqrt{t_b(1-t_b)}\bigr\}$.
Hence its trace‐norm is the sum of the absolute values of those eigenvalues

\begin{equation}
    \begin{aligned}
\big\|\rho^{T_B}\big\|_1
=\mathcal{N'}^{-1}[\;t_b\;+\;\lambda^2(1-t_b)\;+\;2\,\lambda\sqrt{t_b(1-t_b)}]=1+2\dfrac{\lambda\sqrt{t_b(1-t_b)}}{\lambda^2(1-t_b)+t_b}.
    \end{aligned}
    \label{eq:trace‐norm}
\end{equation}
The log‐negativity is then
\begin{equation}
\mathcal{E}_N=\log_2\bigg(1+2\dfrac{\lambda\sqrt{t_b(1-t_b)}}{\lambda^2(1-t_b)+t_b}\bigg)
    \label{eq:LogNegApend}
\end{equation}

\begin{figure}
	\centering
	\includegraphics[width=0.7\linewidth]{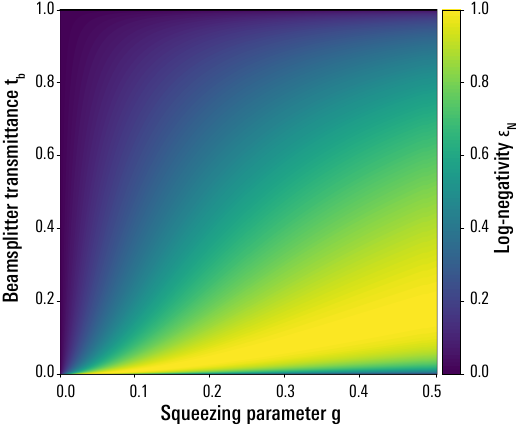}
    \caption{Log-negativity $\mathcal{E}_N$ of the heralded state as a function of the squeezing parameter $g$ and Bob's BS transmittance $t_b$. The colour scale ranges from $\mathcal{E}_N = 0$ (separable) to 1 (maximally entangled), demonstrating the protocol’s tunable entanglement.}
	\label{fig:LogNeg}
\end{figure}

 $\mathcal{E}_N$ is an entanglement monotone that vanishes if and only if the state is separable and provides an upper bound on the distillable entanglement \cite{vidal2002computable,plenio2005logarithmic}.  This behavior is illustrated in Fig.~\ref{fig:LogNeg}, where the log‑negativity is plotted as a function of the tuning parameters $g$ and $t_b$, showing a smooth transition from $\mathcal{E}_N=1$ (maximally entangled) down to $\mathcal{E}_N=0$ (fully separable).

 \section*{IV. Nonlocality Test for Photon-Number Path-Entanglement}
\label{app:Nonlocality Test}

\subsection*{A. Ideal Case}
Here we derive the CH and CHSH inequalities for photon-number path-entangled (PNPE) states probed using displacement (or weak-field homodyne) measurements, assuming the ideal case with no loss. In practice, the displacement is implemented by mixing the mode with a weak local oscillator on a highly transmissive beamsplitter—an approximation valid in the limit of large coherent state amplitude and vanishing beamsplitter reflectivity \cite{banaszek1999testing}.

After Alice and Bob interfere their local modes ($a_1$ and $b_1$) with their respective local oscillators, the joint state becomes
\begin{equation}
  \ket{\psi}_f=\hat{D}_{a_2}(-\delta_\alpha)\hat{D}_{b_2}(-\delta_\beta)\ket{\psi}_{AB},
  \label{eq:final_joint_state}
\end{equation}
where $\hat{D}(-\delta_{\alpha,\beta})$ are displacement operators with $\delta_\alpha = \alpha \sqrt{r_a}$ and $\delta_\beta = \beta \sqrt{r_b}$.
Since we are working with PNPE states, photon counting must be dichotomized into binary outcomes. Here, we group the results into zero and non-zero clicks. This type of measurement encompasses all possible outcomes, including losses, allowing both CH and CHSH inequalities to be applied without relying on assumptions such as fair sampling. In what follows, we focus on the CHSH (Clauser-Horne-Shimony-Holt) inequality, defined as \cite{clauser1969proposed}

\begin{equation}
S=E(\delta_{\alpha_1},\delta_{\beta_1})+E(\delta_{\alpha_1},\delta_{\beta_2})+E(\delta_{\alpha_2},\delta_{\beta_1})-E(\delta_{\alpha_2},\delta_{\beta_2}).
\label{eq:inequity}
\end{equation}

This means we need to calculate the correlation coefficients, $E(\delta_\alpha,\delta_\beta)$, for different settings of $\delta_\alpha\, \&\, \delta_\beta$ to calculate the CHSH parameter S. The correlation coefficient for click or no-click outcomes is given by 
\begin{equation}
\begin{aligned}
E(\delta_\alpha,\delta_\beta)&= P(i=0,j=0|\delta_\alpha,\delta_\beta)+P(i\neq0,j\neq0|\delta_\alpha,\delta_\beta)\\
&-P(i=0,j\neq0|\delta_\alpha,\delta_\beta)-P(i\neq0,j=0|\delta_\alpha,\delta_\beta).
\end{aligned}
    \label{eq:correlation_coef}
\end{equation}

To calculate the probabilities—and consequently the correlation coefficients—we first define the measurement operators. Since we are interested in a binary-outcome measurement (zero vs. non-zero photon counts), this can be described by the following pair of orthogonal projection operators \cite{banaszek1999testing}
\begin{equation}
\hat{\Pi}_0=\ket{0}\bra{0}\;\;,\;\;
\hat{\Pi}_n=\hat{\text{I}}-\hat{\Pi}_0
\label{eq:projection_operators}
\end{equation}

The joint probabilities in Eq.~(\ref{eq:correlation_coef}) can be obtained by simply taking the expectation values of the corresponding operator. Thus, 

\begin{equation}
  \begin{aligned}
  P(i=0,j=0|\delta_\alpha,\delta_\beta)&=\bra{\psi_f}\hat{\Pi}_{0a}\otimes\hat{\Pi}_{0b}\ket{\psi_f}\\
    &=\bra{\psi_{AB}}\hat{D}^\dagger_{b_2}(-\delta_\beta)\hat{D}^\dagger_{a_2}(-\delta_\alpha)\ket{0_{a_2},0_{b_2}}\bra{0_{a_2},0_{b_2}}\hat{D}_{a_2}(-\delta_\alpha)\hat{D}_{b_2}(-\delta_\beta)\ket{\psi_{AB}}\\
&=\braket{\psi_{AB}}{\delta_\alpha,\delta_\beta}\braket{\delta_\alpha,\delta_\beta}{\psi_{AB}}\\
    &=|\braket{\delta_\alpha,\delta_\beta}{\psi_{AB}}|^2,\\
  \end{aligned}
  \label{eq:pnoclick}
\end{equation}

where we used the fact that $\hat{D}^\dagger(-\alpha)=\hat{D}(\alpha)$. 

Note that Eq.~(\ref{eq:pnoclick}) corresponds to projecting the state $\ket{\psi}_{AB}$ onto the coherent state $\ket{\delta_\alpha,\delta_\beta}$, and is, up to a normalization factor, equivalent to the joint Husimi Q-function of the state. Therefore, we write $P(i=0,j=0|\delta_\alpha,\delta_\beta)=Q_{ab}(\delta_\alpha,\delta_\beta)$, and the marginal probability of observing no clicks at one side is given by the marginal Q-function, i.e., $P(i=0|\delta_\alpha)=Q_a(\delta_\alpha)$.

The joint probability of detecting a click at both Alice and Bob is more involved, and is obtained by projecting the state onto the complementary operator $\hat{\Pi}_{na}\otimes\hat{\Pi}_{nb}=\hat{\text{I}}_a\otimes\hat{\text{I}}_b-\hat{\Pi}_{0a}\otimes\hat{\text{I}}_b-\hat{\text{I}}_a\otimes\hat{\Pi}_{0b}+\hat{\Pi}_{0a}\otimes\hat{\Pi}_{0b}$. Hence,
\begin{equation}
  \begin{aligned}
  P(i\neq0,j\neq0|\delta_\alpha,\delta_\beta)&=1-\qdist{\hat{\Pi}_{0a}\otimes\hat{\text{I}}_b} -\qdist{\hat{\text{I}}_a\otimes\hat{\Pi}_{0b}}+\qdist{ \hat{\Pi}_{0a}\otimes\hat{\Pi}_{0b}}\\
  &=1-Q_a(\delta_\alpha)-Q_b(\delta_\beta)_+Q_{ab}(\delta_\alpha,\delta_\beta).
  \end{aligned}
  \label{eq:panyclick}
\end{equation}

The joint probability of no-click at Alice and any-clicks at Bob is
\begin{equation}
  \begin{aligned}
  P(i=0,j\neq0|\delta_\alpha,\delta_\beta)&=\qdist{\hat{\Pi}_{0a}\otimes\hat{\Pi}_{nb}}\\
  &=\qdist{\hat{\Pi}_{0a}\otimes\hat{\text{I}}_b-\hat{\Pi}_{0a}\otimes\hat{\Pi}_{0b}}
  &=Q_a(\delta_\alpha)-Q_{ab}(\delta_\alpha,\delta_\beta).
  \end{aligned}
  \label{eq:panyclick}
\end{equation}

Similarly, one can easily find that $P(i\neq0,j=0|\delta_\alpha,\delta_\beta)=Q_b(\delta_\beta)-Q_{ab}(\delta_\alpha,\delta_\beta)$. 

Substituting these joint probabilities into Eq.~(\ref{eq:inequity}) and Eq.~(\ref{eq:correlation_coef}), we express the CHSH parameter S in terms of the Q-functions:
\begin{equation}
S=2+4[Q_{ab}(\delta_{\alpha_1},\delta_{\beta_1})+Q_{ab}(\delta_{\alpha_1},\delta_{\beta_2})+Q_{ab}(\delta_{\alpha_2},\delta_{\beta_1})-Q_{ab}(\delta_{\alpha_2},\delta_{\beta_2}) -Q_a(\delta_{\alpha_1})-Q_b(\delta_{\beta_1})]
    \label{eq:CHSHinequality}
\end{equation} 

Finally, it is worth mentioning that the CH parameter is related to S via $S=2+4\;CH$.

\subsection*{B. Lossy Case}

In the previous section, we considered the ideal case in which the shared state between Alice and Bob is pure and no losses are present. However, in our proposed implementation, the shared state is described by the mixed density matrix given in Eq.~(\ref{eq:app_initialstates}), and both local losses and non-ideal detectors must be taken into account. We model local losses on Alice and Bob’s sides by introducing fictitious beamsplitters with transmission efficiency $\eta_D$ in front of each detector. In the presence of loss, the POVM elements corresponding to displacement-based click/no-click measurements are given by

\begin{subequations}
\begin{align}
\hat{\Pi}^{(\alpha,\eta_D)}_0&=D^\dagger(\alpha)(1-\eta_D)^{a^\dagger a}D(\alpha),\\
\hat{\Pi}^{(\alpha,\eta_D)}_n&=\hat{\text{I}}-\hat{\Pi}^{(\alpha,\eta_D)}_0.
\end{align}
\label{eq:POVM_with_losses}
\end{subequations}
where $\eta_D$ denotes the total detection efficiency, incorporating both transmission, coupling, and detectors losses. In the following, we consider the symmetric case where both Alice and Bob have have the same detection efficiency.

The joint probability of no-detections is now giving by 

\begin{equation}
    \begin{aligned}
    Q_{ab}^{\eta_D}(\alpha,\beta)&=\text{Tr}[\rho_{AB}(\hat{\Pi}^{(\alpha,\eta_D)}_0\otimes\hat{\Pi}^{(\beta,\eta_D)}_0)]=e^{-(\alpha^2+\beta^2)\eta_D}\\
    &\times\Bigg[
\frac{ 
  t_b 
  + \lambda^2(1 - t_b)\bigl(1 - \eta_D + \eta_D^2|\alpha|^2\bigr)\bigl(1 - \eta_D +  \eta_D^2|\beta|^2\bigr)
  + 2\,\lambda\,\sqrt{(1 - t_b)\,t_b}\,\alpha\,\beta\,\eta_D^2\,
    \cos\bigl(\varphi_a + \varphi_b - \varphi_\alpha - \varphi_\beta\bigr)
}{
  t_b+\lambda^2\,(1 - t_b) 
}\Bigg]
    \end{aligned}
    \label{eq:app_realisitic_jointQ}
\end{equation}

Similarly, the marginal probabilities can be obtained by tracing over the measurement outcomes of the other party. That, is
\begin{subequations}
    \begin{align}
 Q_a^{\eta_D}(\alpha)&=\text{Tr}[\rho_{AB}(\hat{\Pi}^{(\alpha,\eta_D)}_0\otimes\hat{\text{I}}_b)]=
\Big[\frac{t_b + \lambda^2\,(1 - t_b)\,(1 - \eta_D + \alpha^2 \eta_D^2)}{t_b + \lambda^2\,(1 - t_b)}\Big]e^{-\alpha^2 \eta_D}\\
 Q_b^{\eta_D}(\beta)&=\text{Tr}[\rho_{AB}(\hat{\text{I}}_a\otimes\hat{\Pi}^{(\beta,\eta_D)}_0)]=
\Big[\frac{t_b + \lambda^2\,(1 - t_b)\,(1 - \eta_D + \beta^2 \eta_D^2)}{t_b + \lambda^2\,(1 - t_b)}\Big]e^{-\beta^2 \eta_D}
    \end{align}
    \label{eq:app_realisitic_Qs}
\end{subequations}

Substituting these expressions into Eq.~(\ref{eq:CHSHinequality}), we obtain the final form of the CHSH parameter, which we then maximise numerically for each detection efficiency to produce the analytical results shown in the main text.

For the simulation results, the entire experimental layout with photon-number resolving detection was modelled using Strawberry Fields. Numerical optimisation was performed to determine the optimal state preparation and measurement settings for each detection efficiency. The corresponding CHSH values, shown in Fig.~2 in the main text, were obtained using these optimal settings, which are listed in \Cref{tab:optimal_settings}. 

In the numerical model, the preparation of Alice's and Bob's local state includes higher photon number components, up to a cut-off of the Fock state basis at $n = 6$. This introduces a maximum truncation error of approximately $10^{-5}$, given that we operate in the low-gain regime, with the squeezing parameter limited to a maximum of 0.5. Since all gain values in \Cref{tab:optimal_settings} are below 0.5 the truncation error is even smaller than the conservative estimate above, which explains the excellent agreement observed between our numerical simulations and the analytical model.
\begin{table}
    \caption{Optimal settings for each detection efficiency to maximize the CHSH violation in Strawberry Fields simulations }
    \label{tab:optimal_settings}
    \begin{ruledtabular}
    \begin{tabular}{lllllllll}  
      \#& $\eta_D$&$t_b$ & $g$ & $\alpha_1$ & $\alpha_2$ & $\beta_1$ & $\beta_2$&CHSH \\ \hline
       1& 0.65&0.999 & 0.090 & 0.000 & -0.013 & 0.000 & 0.013&2.000000  \\
       2& 0.67	&0.945	&0.048	&0.001	&-0.128&-0.001	&0.128	&2.000002\\
       3& 0.68&	0.905&	0.132&	0.010&	-0.241&-0.010&0.241&2.000133 \\
       4& 0.70&0.823&	0.226&	0.035&	-0.364&-0.035&	0.364&2.002067 \\
       5& 0.75&0.614&	0.279&	0.097&	-0.501&-0.097&	0.501&2.027202 \\
       6& 0.80&0.181&	0.094&	0.139&	-0.554&-0.139&	0.554&	2.088839 \\
       7& 0.85&0.435&	0.320&	0.162&	-0.574&-0.162&	0.574&	2.186472 \\
       8& 0.90&0.414&	0.364&	0.172&	-0.579&-0.172&	0.579&	2.318223 \\
       9& 0.95&0.296&	0.284&	0.172&	-0.573&-0.172&	0.573&	2.483976 \\
       10& 1.00&0.280&	0.300&	0.165&-0.560&-0.165&0.560&  2.685871 \\
    \end{tabular}
    \end{ruledtabular}
\end{table}

\subsection{C. Resilience to Losses}
\begin{figure}[H]
	\centering	\includegraphics[width=0.7\linewidth]{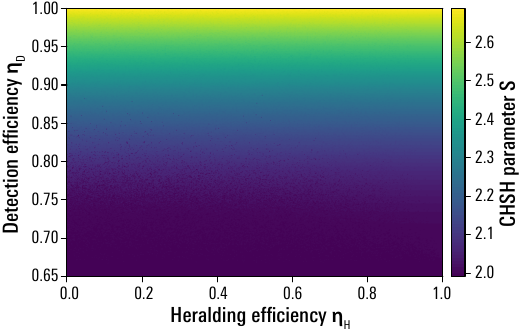}
	\caption{CHSH as function of heralding $\eta_H$ and detection $\eta_D$ efficiencies. We see that our protocol is independent of $\eta_H$. }
	\label{fig:losses_effect}
\end{figure}

Fig.~\ref{fig:losses_effect} shows the CHSH violation as a function of the local detection 
efficiencies at Alice and Bob, and the symmetric heralding efficiency. We see that the violation is almost independent of the losses in the transmitted modes to Charlie, but strongly dependent on the losses in the local mode, which are the losses we wish to minimise to achieve a violation and close the detection loophole.

\section*{\label{app:SuccessRate}V. Success Rate}

Here we derive the success probability for each protocol. In all cases, a successful Bell-state measurement (BSM) occurs when Charlie receives exactly one photon. The probability of this event is given by

\begin{equation}
\begin{aligned}
 P_{\text{succ}}&= \sqrt{\eta_C}(P_A(0)P_B(1)+P_A(1)P_B(0))\\
 &=(1-P_A(1)\sqrt{\eta_C})(P_B(1)\sqrt{\eta_C})+(1-P_B(1)\sqrt{\eta_C})(P_A(1)\sqrt{\eta_C})\\
 &=\sqrt{\eta_C}(P_A(1)+P_B(1))-2\eta_C P_A(1)P_B(1)
\end{aligned}
\label{eq:app_psuccess}
\end{equation}

where $\sqrt{\eta_C}$ is the channel transmittance from Alice or Bob to Charlie, and $P_x(y) $is the probability that the initial state prepared by party $x$ contains $y$ photons in the transmitted mode.

For the protocol proposed in \cite{mycroft2023proposal}, we have $P_A(1) = P_B(1) = \frac{\lambda^2}{1 + \lambda^2}$. In the protocol of \cite{steffinlongo2024long}, assuming ideal heralding detectors, the corresponding probabilities are $P_A(1) = P_B(1) = \frac{\lambda^2}{1 + \lambda^2} t$, where $t$ is the transmission probability through the beamsplitters used to generate the single-photon entangled states. Following the condition $t \ll 1$ assumed in the original work, we set $t = 0.1$.

For our protocol, we have $P_A(1) = \frac{\lambda^2}{1 + \lambda^2}$ and $P_B(1) = \frac{\lambda^2}{1 + \lambda^2} t_b$, where we set $t_b = 0.8$ to maximise the CHSH violation at detection efficiency $\eta_D = 0.75$. We note that in our scheme, one can trade off some Bell violation to increase the success probability. In the limit $t_b \rightarrow 1$, our protocol’s success probability becomes identical to that of the TMSV-based protocol in \cite{mycroft2023proposal}.

\section*{\label{app:asymmetry_analysis}VI. Source of Improvement with Exact Model}

In this section, we investigate the origin of the performance advantage in our protocol and the reason for the differing behaviour exhibited by the states used in various schemes. To this end, we derive an analytical expression for the CHSH inequality as a function of the measurement settings and detection efficiency, assuming systematic loss, for the states
\begin{subequations}
\begin{align}
\ket{\psi}&=c_0\ket{0,1}_{a,b} +c_1 e^{i\varphi}\ket{1,0}_{a,b} \\
\ket{\phi}&=c_0\ket{0,0}_{a,b} +c_1 e^{i\varphi}\ket{1,1}_{a,b}.
\end{align}
\label{eq:two_states}
\end{subequations}

Without loss of generality, the states are expressed in this form to simplify the analysis and to isolate the effect of loss on each Fock component individually. Since the Hilbert spaces for these states are spanned by vacuum and single-photon only (i.e., $\{\ket{0},\ket{1}\}$), then the matrix form of the no-click POVM is given by \cite{vivoli2015comparing}
\begin{equation}
\hat{\Pi}^{(\alpha,\eta)}_0=
\begin{bmatrix}
e^{-\eta|\alpha|^2} & -\eta \alpha^*e^{-\eta|\alpha|^2}  \\
-\eta \alpha e^{-\eta|\alpha|^2}  & (1-\eta+\eta^2|\alpha|^2)e^{-\eta|\alpha|^2}
\end{bmatrix}
\label{eq:POVM_matrix}
\end{equation}

We follow the same procedure outlined in Sec. IV to compute the exact Q-functions for both states and evaluate the CHSH inequality as a function of loss.  This analysis reveals the fundamental difference between the two states in terms of their ability to reach the Eberhard limit.

\subsection*{A. Exact Probabilities for $\ket{\psi}$ with Loss}

The single probabilities of no-detection are given by \cite{banaszek1999testing}
\begin{subequations}
\begin{align}
Q_a^{\psi}(\alpha)&=\qdist{\hat{\Pi}^{(\alpha,\eta)}_0\otimes\hat{\text{I}}_b}= \left[c_1^2 \left(1 - \eta + \eta^2 \alpha^2\right) + c_0^2\right] e^{-\eta |\alpha|^2}\\
Q_b^{\psi}(\beta)&=\qdist{\hat{\text{I}}_a\otimes\hat{\Pi}^{(\beta,\eta)}_0}= \left[c_0^2 \left(1 - \eta + \eta^2 \beta^2\right) + c_1^2\right] e^{-\eta  |\beta|^2}.
\end{align}
\label{eq:probabilities_1}
\end{subequations}

The joint probability of no-detection is 
\begin{equation}
\begin{aligned}
 Q_{ab}^{\psi}(\alpha,\beta)&=\qdist{\hat{\Pi}^{(\alpha,\eta)}_0\otimes\hat{\Pi}^{(\beta,\eta)}_0}\\&= \Big[c_0^2\left(1 - \eta + \eta^2 |\beta|^2\right)+
c_1^2 \left(1 - \eta + \eta^2 |\alpha|^2\right)+ 2 \alpha \beta c_0 c_1 \eta^2 \cos\left(\varphi + \varphi_\alpha - \varphi_\beta\right)\Big]
e^{-\eta( |\alpha|^2+|\beta|^2)}   
\end{aligned}
\end{equation}

\subsection*{B. Exact Probabilities for $\ket{\phi}$ with Loss}

The single probabilities are given by
\begin{subequations}
\begin{align}
Q_a^{\phi}(\alpha)&=\qdist{\hat{\Pi}^{(\alpha,\eta)}_0\otimes\boldsymbol{1}_b}= \left[c_0^2+c_1^2 \left(1 - \eta + \eta^2 \alpha^2\right) \right] e^{-\eta |\alpha|^2}\\
Q_b^{\phi}(\beta)&=\qdist{\boldsymbol{1}_a\otimes\hat{\Pi}^{(\beta,\eta)}_0}= \left[c_0^2+ c_1^2\left(1 - \eta + \eta^2 \beta^2\right)\right] e^{-\eta  |\beta|^2}.
\end{align}
\label{eq:probabilities_1}
\end{subequations}

The joint probability of no-detection is 
\begin{equation}
\begin{aligned}
 Q_{ab}^{\phi}(\alpha,\beta) &= \qdist{\hat{\Pi}^{(\alpha,\eta)}_0\otimes\hat{\Pi}^{(\beta,\eta)}_0} \\
&= \Big[ c_0^2 +
c_1^2 \left(1 - \eta + \eta^2 |\alpha|^2\right)\left(1 - \eta + \eta^2 |\beta|^2\right)  + 2 \alpha \beta c_0 c_1 \eta^2 \cos\left(\varphi - \varphi_\alpha - \varphi_\beta\right) \Big]
e^{-\eta( |\alpha|^2+|\beta|^2)}
\end{aligned}
\end{equation}

\subsection*{C. Observations}
Finally, this analysis reveals the underlying reason for the contrast in performance between the two states. In the Q-function expressions, we observe that loss affects only the coefficient $c_1$ in the state $\ket{\phi}$, whereas in the state $\ket{\psi}$, both $c_0$ and $c_1$ are impacted. This is intuitive: photon loss does not affect the vacuum, but it does degrade the single-photon components of the state. In $\ket{\phi}$, only the $\ket{1,1}$ term is susceptible to loss, while in $\ket{\psi}$, both terms involve single-photon contributions. As a result, minimising $c_1$ in $\ket{\phi}$ can suppress the effect of loss, enabling a Bell violation even under high-loss conditions when optimising the measurement settings, as demonstrated in the main text. This highlights the vacuum as a valuable resource in our protocol.

\end{document}